\documentclass{article}

\usepackage{spconf, amsmath, graphicx}

\usepackage{placeins}   
\usepackage{cleveref}   
\usepackage{amsmath}    
\usepackage{amssymb}    
\usepackage{amsfonts}   
\usepackage{nicefrac}   

\usepackage{graphicx}   
\usepackage{booktabs}   
\usepackage{multirow}   

\usepackage{algorithm}
\usepackage[noend]{algpseudocode}

\usepackage[utf8]{inputenc} 
\usepackage[T1]{fontenc}    
\usepackage{url}            
\usepackage{enumitem}  
\usepackage{xcolor}

\begin{document}
%

\title{FinZero: Launching Multi-modal Financial Time Series Forecast with Large Reasoning Model}

\name{ 
  \shortstack[c]{
    Yanlong Wang$^{1,2,\star}$\thanks{$\star$ Work done during intern at Ant Group.}, Jian Xu$^{1}$, Fei Ma$^{3}$, Hongkang Zhang$^{1}$, Hang Yu$^{4}$, Tiantian Gao$^{1}$, Yu Wang$^{2,6}$ \\ 
    Haochen You$^{5}$, Shao-Lun Huang$^{1\dagger}$, Danny Dongning Sun$^{2\dagger}$, Xiao-Ping Zhang$^{1\dagger}$\thanks{$\dagger$ Corresponding author.}
  }
}

\address{ 
  $^{1}$ Tsinghua University\quad
  $^{2}$ Pengcheng Laboratory\quad
  $^{3}$ Guangming Laboratory\quad
  $^{4}$ Ant Group\\ 
  $^{5}$ Columbia University\quad
  $^{6}$ Southern University of Science and Technology
}

\maketitle

\begin{abstract}
Financial time series forecasting is both highly significant and challenging. Previous approaches typically standardized time series data before feeding it into forecasting models, but this encoding process inherently leads to a loss of important information. Moreover, past time series models generally require fixed numbers of variables or lookback window lengths, which further limits the scalability of time series forecasting. Besides, the interpretability and the uncertainty in forecasting remain areas requiring further research, as these factors directly impact the reliability and practical value of predictions. To address these issues, we first construct a diverse financial image-text dataset (FVLDB) and develop the Uncertainty-adjusted Group Relative Policy Optimization (UARPO) method to enable the model not only output predictions but also analyze the uncertainty of those predictions. We then proposed FinZero, a multimodal pre-trained model finetuned by UARPO to perform reasoning, prediction, and analytical understanding on the FVLDB financial time series. Extensive experiments validate that FinZero exhibits strong adaptability and scalability. After fine-tuning with UARPO, FinZero achieves an approximate 13.48\% improvement in prediction accuracy over GPT-4o in the high-confidence group, demonstrating the effectiveness of reinforcement learning fine-tuning in multimodal large model, including in financial time series forecasting tasks.
\end{abstract}

\section{Introduction}
The field of time series forecasting has garnered increasing attention \cite{wu2021autoformer,luo2024moderntcn,huang2024generative}, as time-series data is widely present in various real-world industries (e.g., transportation, weather, power, finance, etc.). Extracting future trends from historical time-series information holds significant practical value. Among these, financial time series exhibit more distinctive characteristics as they are influenced by more complex factors\cite{wu2021time, baltussen2019indexing}; the asset price movements are shaped by a broad range of external macro- and micro-level influences\cite{lo1988stock, boyer2006crises}, as well as the interplay between buyers and sellers in determining transaction prices\cite{narayanasamy2023relations, malkiel2003efficient, bali2016risk}. This implies that, in such a game-theoretic environment, any discernible patterns or identifiable features (e.g., the pronounced periodicity seen in transportation or power time series) tend to diminish once traders recognize and exploit them for profit. This "adaptive" nature of markets leads to the inability of historical patterns to fully replicate in the future. Predicting such time series is undoubtedly highly challenging. However, even marginal improvements in forecasting performance can yield substantial impacts, particularly in high-frequency trading scenarios\cite{brogaard2023machine,leippold2022machine,gu2020empirical}.

\begin{figure*}[ht]
    \centering
    \vspace{-1ex}
    \includegraphics[width=1.0\textwidth]{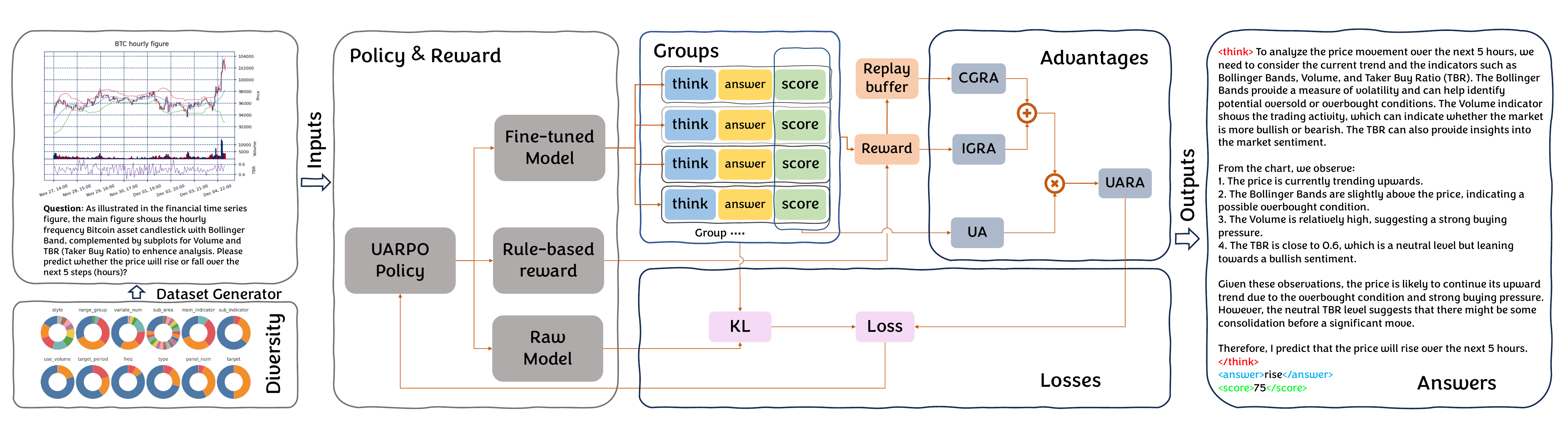}
    \vspace{-3ex}
    \caption{The overall pipeline of the FinZero model, fine-tuned via the UARPO method on the FVLDB.
}
    \vspace{-1ex}
    \label{fig:overall_pipeline}
\end{figure*}

To improve time-series forecasting performance, including financial time series such as exchange rate prediction, specialized models have been designed \cite{lee2024soft,cao2024tempo,wu2022timesnet}. However, several challenges remain unresolved. \textbf{First}, most current time series models require standardization to transform the data into a numerical range that the model can process, such as the normalization techniques used in RevIN\cite{kim2022reversible}. This inevitably leads to the loss of partial information from the original values. \textbf{Second}, patch-based processing is commonly adopted, but this may not fully align with the size and location of critical features\cite{nie2023timeseriesworth64}. \textbf{Third}, time series models usually operate with several fixed configurations, such as lookback window size, variable types and quantities, and data frequency\cite{liu2023itransformer,wu2022timesnet}; these significantly limit their generalizability. Although some pre-trained time-series models\cite{darlow2024dam,tan2024language,zhou2023one,jin2023time} have partially addressed this issue, the advanced reasoning capabilities of large models have not yet been fully leveraged in time series applications. \textbf{Besides}, interpretability of reasoning and uncertainty quantification in forecasting results remain critical yet understudied challenges.

To address the aforementioned challenges, we have abandoned traditional model architectures that process raw time series values and instead transformed the original time series into image compositions. Leveraging reinforcement learning fine-tuning development\cite{guo2025deepseek, achiam2023gpt, bai2022constitutional, xie2025logic, chen2025vinci, peng2025lmm, meng2025mm}, we enhance the visual reasoning capabilities of multimodal large model (MLM). Our focus is on financial time-series trend prediction and reasoning tasks. To support this, we construct the FVLDB dataset, comprising over 10,000 financial time series image-text pairs. To ensure dataset diversity, we performed stratified sampling across multiple dimensions, including asset types, prediction task categories, historical sequence lengths and frequencies, time-series indicator varieties, and image styles.
To tackle the inherent uncertainty and non-stationarity in financial time-series forecasting, we propose the Uncertainty-Adjusted Relative Policy Optimization (UARPO) method. UARPO evaluates both intra-group relative advantage (IGRA) (performance within a group) and cross-group relative advantage (CGRA) (performance between groups over a recent window). Additionally, it adjusts advantage levels based on prediction uncertainty (Uncertainty-Adjusted Relative Advantage, UARA).

In this work, we propose the FinZero model, as illustrated in \Cref{fig:overall_pipeline}, which fine-tunes 3B-parameter multimodal large model via the UARPO method in the FVLDB dataset, which enables MLM to explicitly account for prediction uncertainty. Comparative experiments with GPT-4 show a 13.48\% improvement in prediction accuracy in the high-confidence group, validating the effectiveness of RL-based cross-modal fine-tuning for financial time-series forecasting and reasoning. By providing confidence score and reasoning traces, FinZero helps users better understand model predictions and their rationale, ultimately supporting more informed financial decision-making, making it particularly valuable for real-world financial applications where risk assessment is paramount.

\vspace{-1ex}  
\section{Methods}
\subsection{Uncertainty Adjusted Related Policy Optimization}
The GRPO\cite{shao2024grpo} is employed to fine-tune the DeepSeek-R1\cite{deepseekai2025deepseekr1}. As an improvement over the PPO\cite{schulman2017ppo}, GRPO eliminates the need for an additional model as a policy model (as required by methods like PPO) and leverages Group Relative Advantage sampled from multiple outputs within a group, thereby avoiding the necessity for extra value function approximation. GRPO primarily focuses on the relative advantages among multiple outputs within each sample group, while other methods like REINFORCE++\cite{hu2025reinforcepp} utilize discounted cumulative rewards to construct advantage variations that reflect the training process, which helps improve training stability. Additionally, how to reflect the uncertainty in model inference results holds significant importance, as it aids decision-making by assessing the confidence level of reasoning outcomes.

Based on the above, we propose the UARPO algorithm, which introduces two key enhancements. 1.Under the same prediction target, a multidimensional advantage function combining In-Group Relative Advantage (IGRA) within samples and Cross-Group Relative Advantage (CGRA) across groups; 2.Construction of an uncertainty function (UA) based on the model’s inference confidence scores, ultimately forming Uncertainty-Adjusted Relative Advantage (UARA). The optimization objective can be expressed as Equation \ref{eq:uarpo_obj}

\begin{equation}
\small  
\begin{split}
J&_{\text{UARPO}}(\theta) = E\left[ q \sim P(Q), \{o_i\}_{i=1}^G \sim \pi_{\theta_{\text{old}}}(O|q), \tau \in \mathcal{T} \right] \\
& \frac{1}{G}\sum_{i = 1}^{G}\frac{1}{|o_i|}\sum_{t = 1}^{|o_i|} \Bigg[\min\left\{\frac{\pi_{\theta}(o_{i,t}|q,o_{i,<t})}{\pi_{\theta_{\text{old}}}(o_{i,t}|q,o_{i,<t})}\left(\hat{A}^I_{i,t}  
+\hat{A}^{S_{\tau}}_{t}\right)\hat{U}_{i,t},\right. \\
& \left.\text{clip}\left(\frac{\pi_{\theta}(o_{i,t} \mid q,o_{i,<t})} {\pi_{\theta_{\text{old}}}(o_{i,t} \mid q,o_{i,<t})}, 1 - \varepsilon, 1 + \varepsilon\right)\left(\hat{A}^I_{i,t}+\hat{A}^{S_{\tau}}_{t}\right)\hat{U}_{i,t}  \right\} \\
&- \beta D_{\text{KL}}\left[\pi_{\theta}||\pi_{\text{ref}}\right]\Bigg]
\end{split}
\label{eq:uarpo_obj}
\end{equation}

\begin{equation}
\hat{A}^I_{i,t} \triangleq \widetilde{r}_i = \frac{r_i - \text{mean}(\mathbf{r})}{\text{std}(\mathbf{r})}
\end{equation}

\begin{equation}
\hat{A}^{S_{\tau}}_{t} \triangleq \widetilde{s}_t^{\tau} = \frac{s_t^{\tau} - \text{mean}(\mathbf{s^{\tau}_{t-l,t}})}{\text{std}(\mathbf{s^{\tau}_{t-l,t}})} 
\end{equation}

Where $\pi_\theta$ and $\pi_{\theta_{old}}$ are the current and old policy models, $q$ and $o$ are questions and outputs sampled from the question dataset and the old policy $\pi_{\theta_{old}}$, respectively. $\varepsilon$ is a clipping-related hyper-parameter introduced in PPO for stabilizing training. $\hat{A}^I_{i,t}$ represents the in-group relative advantage as in GRPO, where $\mathbf{r}=[r_0,r_1, \dots,r_i,\dots,r_G ]$, $\hat{A}^{S_{\tau}}_{t}$ represents the cross-group relative advantage where $\mathbf{s^{\tau}_{t-l,t}}=[s_{t-l},s_{t-l+1}, \dots,s_t| \tau]$ and $s_t^{\tau}=\frac{1}{G}\sum_{i=1}^Gr^{\tau}_{i,t}$, which indicate the advantage of the current group's overall performance relative to the average performance over multiple steps in a recent window period under the same prediction objective. $\mathbf{s^{\tau}_{t-l,t}}$ is a group consisting of multiple steps with window length $l$. $\hat{U}_{i,t}\triangleq \alpha \cdot\frac{\text{score}-\text{const}}{100}$ is the uncertainty adjustment function, and $\alpha$ denotes an adjustable coefficient. The algorithmic iterative process can be described as \Cref{algo:uarpo}.

\begin{algorithm}[t]
\scriptsize
\caption{Iterative UARPO}
\label{algo:uarpo}
\begin{algorithmic}[1]
\State \textbf{Input}: Initial policy model $\pi_{\theta_{\text{init}}}$; reward model $r_\phi$; task prompts $\mathcal{D}$; hyperparameters $\epsilon, \beta, \mu$; stack length $L$
\State \textbf{Initialize}: policy model $\pi_\theta \leftarrow \pi_{\theta_{\text{init}}}$; target special stack $\mathbf{s^{\tau}_L}$

\For{iteration $= 1$ \textbf{to} $I$}
    \State Update reference model $\pi_{\text{ref}} \leftarrow \pi_\theta$
    \State Initialize stack $\mathcal{S}[0..L-1]$ 
    
    \For{step $= 1$ \textbf{to} $M$}
        \State Sample batch $\mathcal{D}_b \subset \mathcal{D}$
        \State Update the old policy $\pi_{\theta_{\text{old}}} \leftarrow \pi_\theta$
        \State Sample $G$ outputs $\{o_i\}^G_{i=1} \sim \pi_{\theta_{\text{old}}}(\cdot|q)$ for each question $q \in \mathcal{D}_b$.
        \State Compute rewards $\{r_i\}^G_{i=1}$ and confidence scores $\{u_i\}^G_{i=1}$ for each output $O_i$ by runnning $r_\phi$.
        \State Compute current step average reward $\frac{1}{G}\sum^G_{i=1}r^{\tau}_{L}$ for current target $\tau$.
        \State Compute $\hat{A}^I_{i,t}$ for the $t$-th token of $o_i$ through group relative advantage estimation.
        \If{$step > L$}
        \State Compute $\hat{A}^I_{i,t}$ for the $t$-th token through latest L step relative advantage estimation for target $\tau$
        \EndIf
        \State Gather two part relative advantage and multiply with coressponding confidence score
        \For{UARPO iteration $=1,...,\mu$}
            \State Update the policy model $\pi_{\theta}$ by maximizing the UARPO objective.
        \EndFor
        \State Update $r_{\phi}$ through continuous traning using a replay mechanism.
    \EndFor
    
    \State Update $r_\phi$ with replay mechanism
\EndFor
\State \textbf{Output}: $\pi_\theta$
\end{algorithmic}
\end{algorithm}

\subsection{Rewards and Uncertainty}

\begin{itemize}[leftmargin=*]
\item \textbf{Accuracy Reward} Prediction accuracy is commonly used to evaluate the performance of reinforcement learning models and construct loss functions. Specifically, it measures the consistency between the model’s predictions and the ground-truth outcomes (rise/fall) of each sample.

\item \textbf{Completion Length Reward} Previous works have found that text length expansion occurs in large model RL reasoning, which is helpful for improving reasoning time and enabling complex reasoning. Therefore, we provide this type of reward. Specifically, when the text reasoning length is no more than 200 tokens, a gradually increasing reward is offered.

\item \textbf{Format Reward} Format reward help the model learn the target output format during fine-tuning.

\item \textbf{Confidence Score} Prior works (\cite{emnlp24liu, uncer24xia}) have explored the feasibility and methods for large models to learn task uncertainty. Given the high uncertainty inherent in financial decision-making—where uncertainty analysis is critical for model development and real-world use—we integrate model reasoning uncertainty into reinforcement learning fine-tuning. During each image-text reasoning process, the model outputs a confidence score based on the input information and its reasoning. This score quantifies the model’s uncertainty about its reasoning result for the given task, enabling it to learn problem difficulty and uncertainty through training.
\end{itemize}

\vspace{-1ex} 
\section{Experiments}
\label{sec:experiments}

\subsection{FVLDB Dataset}

\begin{figure*}[ht]
    \centering
    \includegraphics[width=0.9\textwidth]{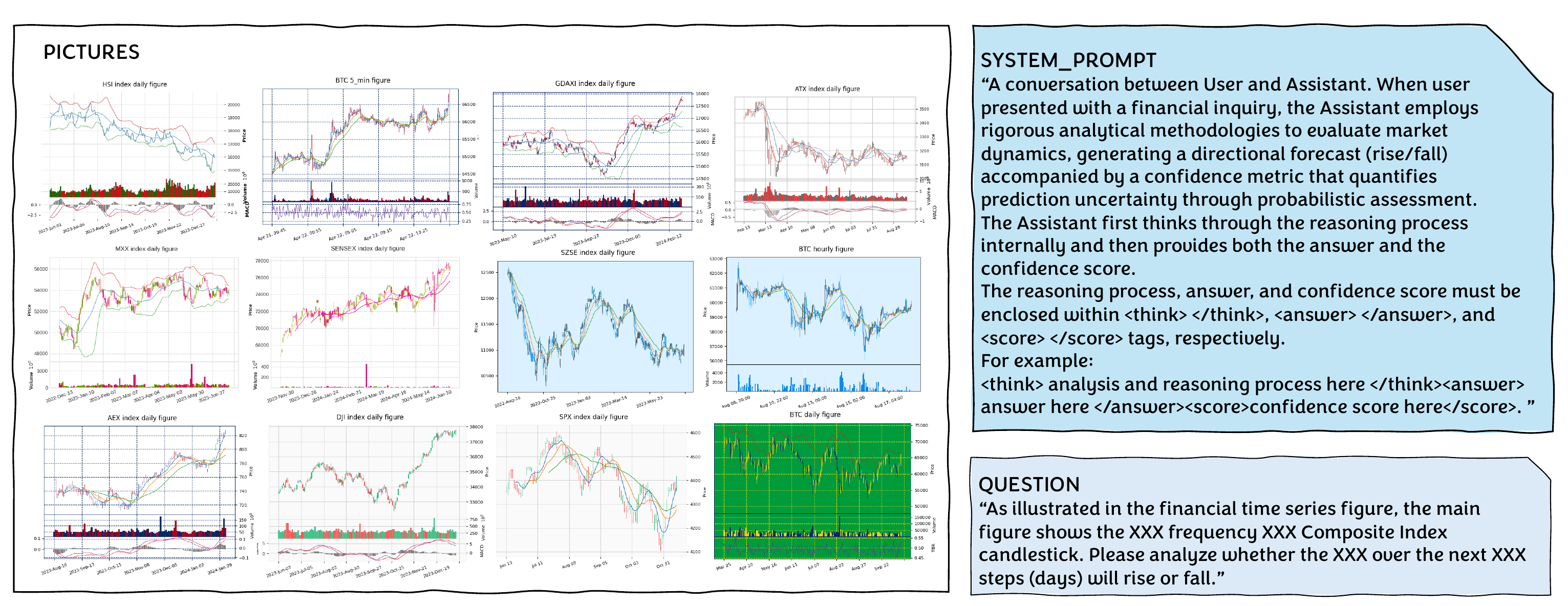}
    \vspace{-1ex}
    \caption{Overview of Image-Text Pairs for the FVLDB Dataset.
}
    \label{fig:dataset_pic}
    \vspace{-1ex}
\end{figure*}

To validate our idea, we specifically construct a financial time-series image-text dataset (FVLDB as \Cref{fig:dataset_pic}) with over 10000+ samples. The images in FVLDB contain a wealth of financial assets, along with corresponding text descriptions and questions. To enhance data diversity, FVLDB includes index data from global stock markets, as well as data on cryptocurrency assets such as Bitcoin. The time-series length, sampling frequency, type, and number of features of the assets in each image are variable, and the image styles are also diverse. This flexibility enables the model to process diverse data types.

\begin{figure}
    \centering
    \includegraphics[width=\columnwidth]{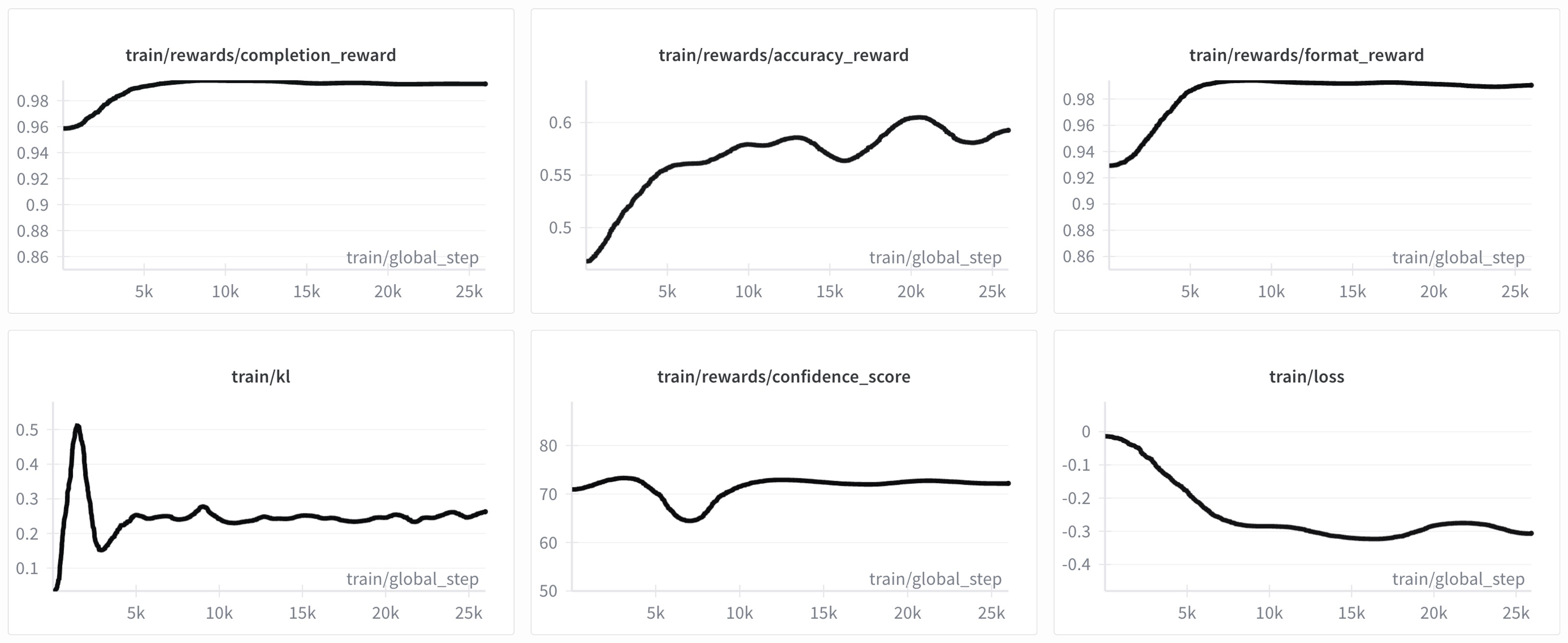}
    \vspace{-2ex}
    \caption{Overview of the FinZero Training.}
    \label{fig:training_process}
    \vspace{-1ex}
\end{figure}

\vspace{-1ex} 
\subsection{Setup}
The FinZero utilize the Qwen2.5-VL-3B model as the backbone and fine-tuned it directly on the FVLDB dataset with the UARPO algorithm. For baselines, we select the original Qwen2.5-VL-3B model, the Qwen2.5-VL-7B model, and the larger-scale GPT-4o. Additionally, we also fine-tuned Qwen2.5-VL-3B with GRPO, and also constructed a Naive Model, which extends the trend of the past period of time. The Adam optimizer was adopted with a learning rate of 1e-6, and the fine-tuning process ran for two epochs. All experiments were conducted on a server equipped with two 80G Nvidia A100 GPUs.
\subsection{Results}

\begin{table}
    \centering
    \small
    \caption{Main Results of Model Accuracy Comparison.}
    \label{tab:main_results}
    \resizebox{\linewidth}{!}{
    \begin{tabular}{c|c|c|c|c|c|c|c|c}
    \midrule
        \textbf{} & \multicolumn{4}{c|}{\textbf{Volitality ACC (\%)}} & \multicolumn{4}{c}{\textbf{Price ACC (\%)}} \\ \midrule
        \textbf{Model} & 5 & 21 & 63 & Avg & 5 & 21 & 63 & Avg \\ \midrule
        \textbf{Naive} & 48.54 & 46.23 & 48.46 & 47.75 & 50.00 & 52.04 & 50.00 & 50.68 \\ \midrule
        \textbf{Qwen2.5-VL-3B} & 46.67 & 45.69 & 50.51 & 47.62 & 54.20 & 51.64 & 52.54 & 52.79 \\ \midrule
        \textbf{Qwen2.5-VL-7B} & 50.49 & 43.64 & 51.16 & 48.43 & \underline{55.55} & 51.91 & 51.14 & 53.53 \\ \midrule
        \textbf{GRPO} & 53.68 & \underline{54.86} & 52.15 & \underline{53.56} & 53.24 & \underline{53.63} & \underline{53.76} & \underline{53.54} \\ \midrule
        \textbf{GPT-4o} & \underline{54.28} & 48.26 & \textbf{53.38} & 51.97 & \textbf{56.16} & 51.22 & 51.14 & 52.84 \\ \midrule
        \textbf{FinZero} & \textbf{56.31} & \textbf{65.74} & \underline{52.93} & \textbf{58.33} & 54.52 & \textbf{56.31} & \textbf{65.88} & \textbf{58.90} \\ \midrule
    \end{tabular}
    }
\end{table}

\begin{table}
    \centering
    \scriptsize
    \caption{Model Prediction Accuracy Across Confidence Score Groups.}
    \label{tab:uncertainty_group}
    \resizebox{0.8\linewidth}{!}{
    \begin{tabular}{c|c|c|c}
        \midrule
        \textbf{} & \textbf{Low (\%)} & \textbf{Middle (\%)} & \textbf{High (\%)} \\ \midrule
        \textbf{Qwen2.5-VL-3B} & 51.2 & 51.7 & 49.3 \\ \midrule
        \textbf{Qwen2.5-VL-7B} & 47.38 & 47.81 & 54.36 \\ \midrule
        \textbf{GRPO} & 53.85 & 53.19 & 54.61 \\ \midrule
        \textbf{GPT-4o} & 49.85 & 49.42 & 54.75 \\ \midrule
        \textbf{FinZero} & 54.48 & 56.67 & 62.13 \\ \midrule
    \end{tabular}
    }
\end{table}

As shown in \Cref{fig:training_process}, the model’s rewards continuously increase during the UARPO fine-tuning process: the format reward and completion length reward rise rapidly in the early stage of training and then stabilize, while the accuracy reward also increases steadily with training; meanwhile, the loss value decreases consistently. The prediction performance of the fine-tuned model on the test set is presented in \Cref{tab:main_results}. After UARPO fine-tuning, FinZero exhibits more competitive prediction performance compared to baseline models, whether in price prediction tasks or volatility prediction tasks. While FinZero with 3B parameter size surpasses larger parameter models such as GPT-4o. Additionally, when test set samples are divided into three equal groups based on the model’s uncertainty scores sorted from highest to lowest as in \Cref{tab:uncertainty_group}, it shows that for the FinZero, the prediction accuracy of samples with high confidence scores is further improved—the prediction accuracy of the highest-score group is increased by approximately 13.5\% relative to that of GPT-4o. 

\begin{figure}
    \centering  
    \small
    \vspace{-1ex}  
    
    \includegraphics[width=0.7\linewidth]{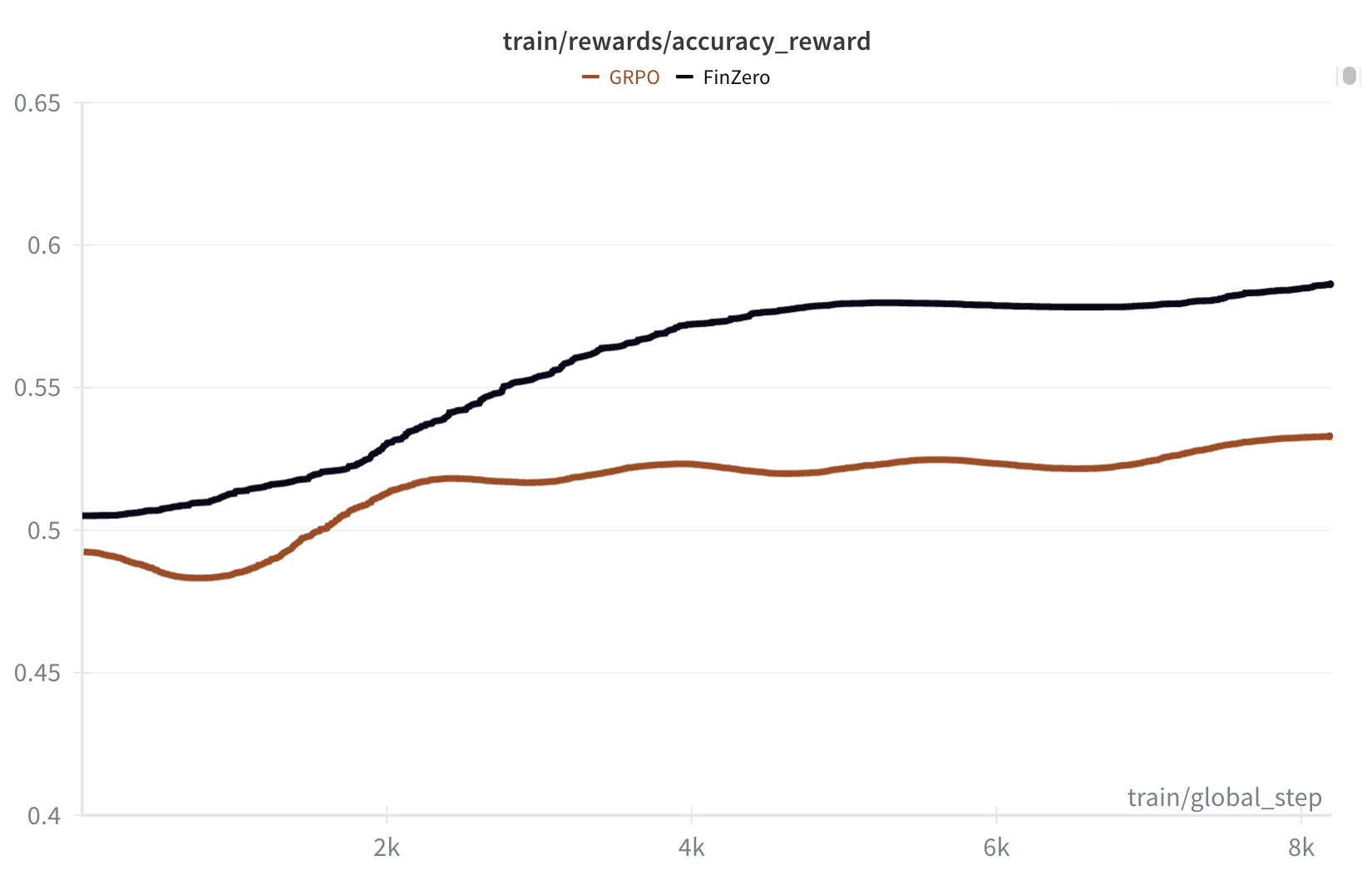}
    \vspace{-1ex}  
    
    \caption{Finetuning Accuracy Trend Comparison.}
    \vspace{-3ex} 
    \label{fig:finetuning_compare}
\end{figure}

Furthermore, compared to Qwen2.5-VL-3B fine-tuned by GRPO, FinZero achieves better average prediction performance. Meanwhile, grouping based on confidence scores exhibits a more pronounced positive correlation with prediction accuracy. Besides, we illustrate the accuracy changes of the two models during the fine-tuning process, as shown in \Cref{fig:finetuning_compare}.

\vspace{-3ex}
\section{Conclusions}
\label{sec:conclusions}
\vspace{-1ex}
This study introduces FinZero, a model for multimodal financial time-series reasoning. The FinZero achieve competitive forecasting accuracy rivaling larger models like Qwen-7B and GPT-4o. Moreover, FinZero provides uncertainty scores that reliably indicate prediction confidence—higher scores correlate with greater accuracy. This work demonstrates the potential of cross-modal reinforcement learning to advance financial reasoning, offering both a robust method and a practical tool for reliable predictions.

\clearpage

{\footnotesize 

\bibliographystyle{IEEEbib}
\bibliography{reference}
}
\end{document}